\documentclass[%
 reprint,
superscriptaddress,
 amsmath,amssymb,
 aps,
floatfix,
]{revtex4-2}

\usepackage[dvipdfmx]{graphicx}
\usepackage{dcolumn}
\usepackage{bm}
\usepackage{braket}
\usepackage{amsmath}
\usepackage{ulem}
\usepackage{xcolor}
\DeclareRobustCommand{\erase}{\bgroup\markoverwith{\textcolor{red}{\rule[.5ex]{2pt}{0.4pt}}}\ULon}

\begin{document}

\preprint{APS/123-QED}

\title{Detection of ferroic octupole ordering in $d$-wave altermagnetic rutile-type compounds}

\author{Masaichiro Mizumaki}
\email{mizumaki@kumamoto-u.ac.jp}
\affiliation{%
Faculty of Science, Kumamoto University, 2-39-1 Kurokami Chuou-ku, Kumamoto 860-8555, Japan
}%

\author{Norimasa Sasabe}
\affiliation{Center for Basic Research on Materials, National Institute for Materials Science, Tsukuba, Ibaraki 305-0047, Japan}%

\author{Takayuki Uozumi}
\affiliation{
Department of Physics and Electronics, Osaka Metropolitan University,
Sakai, Osaka 599-8531, Japan
}%

\author{Rikuto Oiwa}
\affiliation{
Graduate school of Science, Hokkaido University,
Sappro, Hokkaido 060-0810, Japan
}%

\author{Hiroaki Kusunose}
\affiliation{
Department of Physics, Meiji University,
Kawasaki, Kanagawa 214-8571, Japan
}%

\date{\today}

\begin{abstract}
We propose that X-ray absorption and emission magnetic circular dichroism (XAS-MCD and XES-MCD) are promising measurements to directly detect ferroic higher-rank multipoles as candidate order parameters in altermagnets.
Using the sum rules for XES-MCD and connecting them to multipole language, we demonstrate that the expectation value of the magnetic octupole moment is finite in the $d$-wave altermagnetic candidate rutile-type compounds TF$_2$ (T=transition metal).
We also perform spectral calculations of XAS-MCD and XES-MCD based on an effective model with a full multiplet approach.
While the intensity of the XAS-MCD spectra vanishes, the XES-MCD spectra exhibit finite intensity, whose spectrum becomes opposite by inverting the N\`eel vector. These results clearly indicate ferroic magnetic octupole order in these compounds.
\end{abstract}

\maketitle

X-ray magnetic circular dichroism (XMCD), discovered by Sh\"{u}tz $et$ $al.$~\cite{SHutzPRL}, has developed as a powerful element- and orbital-selective measurement for probing the electronic states including spins with broken macroscopic time-reversal symmetry~\cite{ChenPRB1990, Chen1995, Alders1998, vanDerLaan2014}. In particular, in the soft X-ray region, XMCD provides a quantitative tool to independently evaluate the orbital $L_z$ and spin $S_z$ magnetic moments using sum rules~\cite{PCarraSumL92, PCarraSumS93}.

Even in noncollinear antiferromagnets such as Mn$_3$Sn and collinear antiferromagnets such as MnTe~\cite{MazinPRB23,AoyamaPRM24,OsumiPRB24}, a possible observation of XMCD has been theoretically proposed~\cite{YamasakiJPSJ, SasabePRL1, SasabePRL2,HarikiPRL24}, and in the former, experimentally observed~\cite{KimataNC, sakamoto2021observation, sakamoto2024bulk}.
The MCD signal in these antiferromagnets arise from the so-called anisotropic magnetic dipole moment $T_z$ ($\mathbb{M}_z''$ in Table~I) in the sum rules.
$T_z$ is also crucial for understanding the anomalous Hall effect and related phenomena observed in MnTe and Mn$_3$Sn ~\cite{hayami2021essential, OhgataPRB2025}.

In the dipole transition process in XMCD, the measurable quantities are limited to lower-rank multipoles $-$$L_z$, $S_z$, and $T_z$$-$ all rank-1 vectors~\cite{YamasakiarXiv}. However, higher-rank multipoles can dominate certain physical properties. For example, in NpO$_2$~\cite{NPO2TokunagaPRL} and Ce$_x$La$_{1-x}$B$_6$~\cite{MannixPRL, MatsumuraPRL, Kusunose2005JPSJ}, an octupole serves as an order parameter in one of the ordered phases within their rich phase diagrams.
Direct observation of higher-rank multipoles is therefore essential to understanding hidden orders and electronic states in magnetic materials.

In recent years, ``altermagnet'', exemplified by MnTe, has attracted considerable attention, because it breaks global time-reversal symmetry similar to ordinary ferromagnets, even in the antiferromagnetic alignment of magnetic ions placed in anisotropic environments~\cite{Naka2019NatureCom,KyoHoon2019PRB,Smejkal2020MS}.
Altermagnets can also be understood by a ferroic ordering of higher-order multipoles, such as a magnetic octupole~\cite{SpaldinPRX2024,McClarty2024PRL}, which is underlie for the piezomagnetic effect and strong magnetic anisotoropies. 
In this Letter, we propose that the resonant X-ray Emission MCD (RXES-MCD) is a promising probe for the detection of ferroic octupole ordering by establishing a direct link between the coupled tensor $w^{ijk}$ in the sum-rule formula~\cite{Laan1994PRB,Laan1996PRB} and the complete set of multipole operator basis $X_{lm}^{(s)}(k)$~\cite{Kusunose2020JPSJ, Kusunose2024JPSJ}, each possessing clear physical meaning.
We further demonstrate that X-ray emission spectroscopy shows finite intensity in magnetic octupole ordering in TF$_2$ (T=transition metal).

\begin{figure}[t]
  \begin{center}
   \includegraphics[width=80mm]{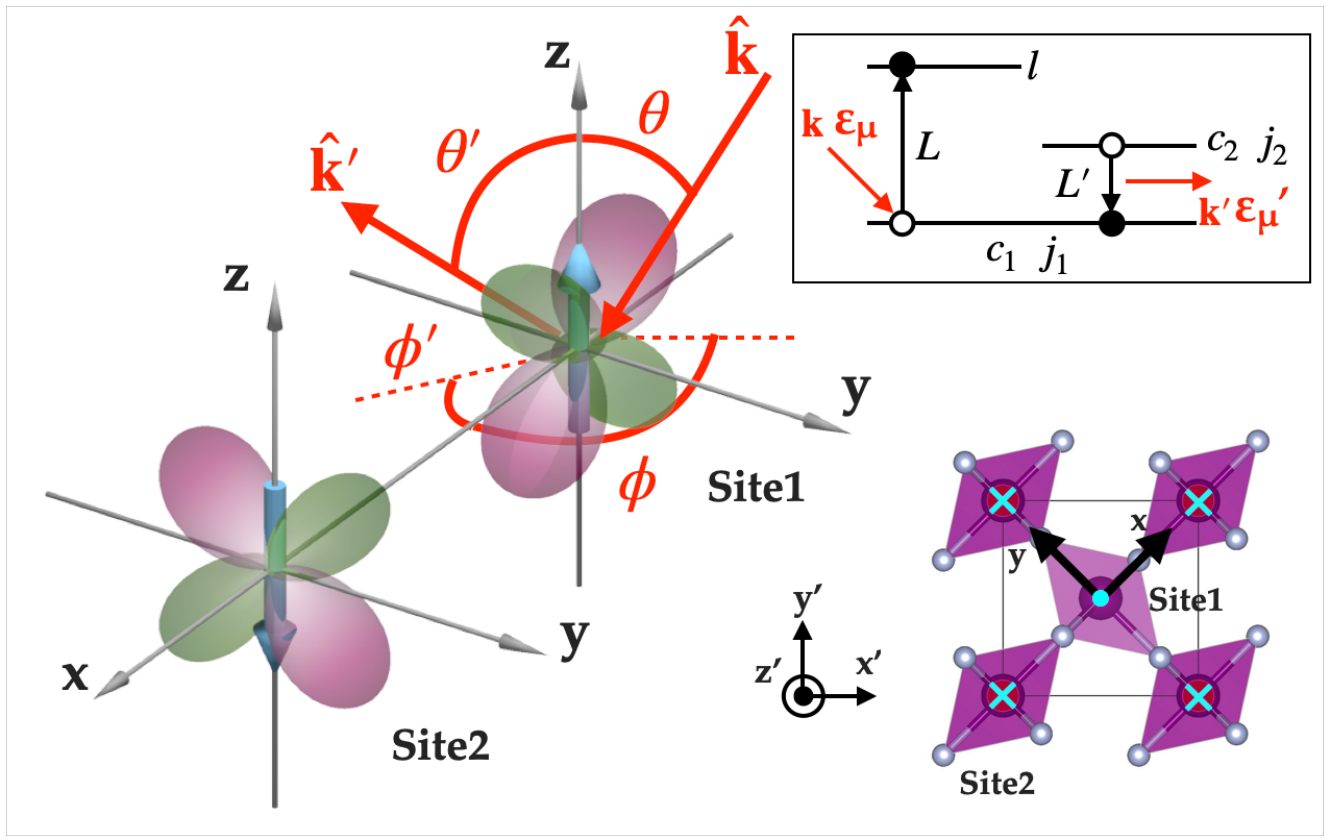}
  \caption{
  (Left) Geometry of the resonant XES process and 3$d$ orbital with spin (Blue arrow) of local coordinate at Site~1 and Site~2 under $D_{2h}$ symmetry.
  (Right) Crystal Structure of rutile. (Inset) One electron process in the resonant XES, where symbols are explained in the main text.}
  \end{center}
\end{figure}

As shown in Fig.~1, the RXES process is described by placing the target element at the origin and choosing the quantization axis along the direction of the magnetic moment.
The incident photon has polarization $\epsilon_{\mu}$, wave vector $\bm{k}$ ($=\lambda^{-1}$) with angles $\hat{\bm{k}}=(\theta,\phi)$, and energy $\hbar\omega_{\bm{k}}$, while the emitted photon has polarization $\epsilon_{\mu'}$, wave vector $\bm{k'}$ [$\hat{\bm{k}}'=(\theta',\phi')$], and energy $\hbar\omega_{\bm{k'}}$. The scattering intensity is expressed using the Kramers-Heisenberg formula,
\begin{eqnarray}
I_{\mu}&=&\sum_{f,\mu'}\left|\sum_m\frac{\bra{f}\bm{\epsilon}_{\mu'} \cdot \bm{r}\ket{m}\bra{m}\bm{\epsilon}_{\mu} \cdot \bm{r}\ket{g}}{\hbar\omega_{\bm{k}}+E_g-E_m+i\Gamma_n/2}\right|^2
\notag\\&\quad&
\times\delta(\hbar\omega_{\bm{k}}+E_g-\hbar\omega_{\bm{k}'}-E_f),
\end{eqnarray}
where $E_g, E_m,$ and $E_f$ denote the energies of the initial state $\ket{g}$, intermediate state $\ket{m}$, and final state $\ket{f}$, respectively.
Within the fast-collision approximation $-$assuming the second-order resonant process occurs so rapidly that excitation and de-excitation effectively occur simultaneously$-$ the coupled tensor expansion was performed~\cite{BorgattiRPB2004, VeenendaalPRB1996}.
When the resonant process shown in (1) is described as a one-electron process, it takes the form illustrated in the inset of Fig.~1: the excitation via an electric $2^L$-pole transition corresponds to a transition from a core level with orbital quantum number $c_1$ and total angular momentum $j_1 = c_1 \pm \tfrac{1}{2}$ to the valence band with orbital quantum number $l$, where $L = |l - c_1|$. Conversely, the de-excitation involves filling the core level $(c_1, j_1)$ through an electric $2^{L'}$-pole decay transition from another core level $(c_2, j_2)$, where $L' = |c_1 - c_2|$.

The sum rule of RXES~\cite{BorgattiRPB2004} intensity is expressed as follows (see Supplemental Material in detail~\cite{test}):
\begin{eqnarray}
I^{\text{RXES}}_{\mu}&=&\mathbb{N}(lLL'\omega)\sum_{\mu'zz'r} (-1)^{z+z'}\left[\frac{z,z'}{r}\right]^{1/2}
\notag\\&\quad&
\times C^{z0}_{L\mu,L\text{-}\mu}C^{z'0}_{L'\mu',L'\text{-}\mu'}
\Theta^{zz'r}\left(\boldsymbol{\hat{k}},\boldsymbol{\hat{k'}}\right)
\notag\\&\quad&
\times \mathbb{E}^{zz'r}_{j_1}\left(c_1lL\right)
\mathcal{\tilde{B}}^{z'}_{j_1,j_2}\left(c_1c_2L'\right),
\end{eqnarray}
where we set $\mu(\mu') = \pm 1$ for MCD, corresponding to right- and left-circular polarization of the incident and emitted photons, respectively.
Here, $\mathbb{E}^{zz'r}_{j_1}\left(c_1lL\right)$ is the term associated with the transition probability of the excitation process:
\begin{eqnarray}
\mathbb{E}^{zz'r}_{j_1}\left(c_1lL\right)=\sum_{ab}\mathcal{\tilde{C}}^{abrzz'}_{j_1}
w^{abr}_{\rho}(l),
\end{eqnarray}
where the coupled tensor $w^{abr}_{\rho}$, containing various degrees of freedoms of the relevant electronic states, is given using
\begin{eqnarray}
w^{abr}_{\rho}(l)=(-1)^{a+b+\rho}\underline{n}_{abr}^{-1}\sum_{\xi,\eta}w^{ab}_{\xi,-\eta}(l)
\begin{pmatrix}
a & r & b\\
-\xi & \rho & \eta\\
\end{pmatrix},
\end{eqnarray}
and $\mathcal{\tilde{C}}^{abrzz'}_{j_1}$ is its coefficient. Furthermore $\underline{n}$ is the normalization factor of $w^{abr}$ and $w^{ab}_{\xi,-\eta}(l)$ is the uncoupled tensor~\cite{VeenendaalPRB1996,test}.
$\mathcal{\tilde{B}}^{z'}_{j_1,j_2}\left(c_1c_2L'\right)$ is the probability of detecting z$'$-polarized radiation emitted in the $j_2\rightarrow j_1$ decay~\cite{VeenendaalPRB1996}.
$\Theta^{zz'r}\left(\boldsymbol{\hat{k}},\boldsymbol{\hat{k'}}\right)$ represents the geometrical configuration of incident and emitted photons, and $C^{z0}_{L\mu,,L\text{-}\mu}$ is the Clebsch-Gordan coefficient.
We have also used the abbreviated notation $[a,\dots,b] = (2a+1)\cdots(2b+1)$, and $\mathbb{N}(lLL'\omega)$ is a global normalization constant, expressed in terms of the reduced scattering amplitude $\mathcal{A}^{L',\lambda'}_{L,\lambda}(\omega_{\bm{k}})$.

\renewcommand{\arraystretch}{1.8}
\begin{table}[t]
 \caption{Correlation between coupled tensor and multipole basis.
 Columns $s,p,$ and $d$ represent the coefficient in (5).
 M(E) represents magnetic(electric). AM means angular momentum.
 Note that $r$ is the rank of multipole, and with(without) prime represents spinful(spinless) multipoles in the column of example. Abbreviations for anisotropy are $u=3z^2-r^2$, $\alpha_z=z(5z^2-3r^2)$, and $4a=35z^4-30z^2r^2-3r^4$.
 }
 \centering
\begin{tabular}{l c c c c c c  }
\hline
\hline
type & $w^{abr}$ & multipole & example & $s$ & $p$ & $d$ \\
\hline
number &  $w^{000}$ & $Q^{(\text{orb})}_{00}\sigma_0$ & $\mathbb{Q}_0$ & 1 & 1 & 1 \\
SOC & $w^{110}$ & $Q^{(1)}_{00}(-1)$ & $\mathbb{Q}_0'$ & - & $\sqrt{3}$ & $\frac{\sqrt{3}}{2}$ \\
\hline
orbital AM& $w^{101}$ & $M^{(\text{orb})}_{1\zeta}\sigma_0$ & $\mathbb{M}_z$ & - & -1 & -$\frac{1}{2}$ \\
spin AM& $w^{011}$ & $M^{(1)}_{1\zeta}(-1)$ & $\mathbb{M}_z'$ & -1 & -1 & -1 \\
AMD& $w^{211}$ & $M^{(1)}_{1\zeta}(+1)$ & $\mathbb{M}_z''$ & - & $\frac{5\sqrt{10}}{2}$ & $\frac{7\sqrt{10}}{4}$  \\
\hline
E quadrupole& $w^{202}$ & $Q^{(\text{orb})}_{2\zeta}\sigma_0$ & $\mathbb{Q}_{u}$ & - & -5 & -$\frac{7}{2}$  \\
\hline
M octupole & $w^{303}$ & $M^{(\text{orb})}_{3\zeta}\sigma_0$ & $\mathbb{M}_{xyz}$ & - & - & $\frac{7}{3}$ \\
M octupole & $w^{213}$ & $M^{(1)}_{3\zeta}(-1)$ & $\mathbb{M}_{xyz}'$, $\mathbb{M}_{\alpha_z}'$ & - & $\frac{5\sqrt{15}}{3}$ & $\frac{7\sqrt{15}}{6}$ \\
\hline
E hexadecapole& $w^{404}$ & $Q^{(\text{orb})}_{4\zeta}\sigma_0$ & $\mathbb{Q}_{4a}$ & - & - & $21$ \\
\hline
\hline
\label{tbl1}
\end{tabular}
\end{table}

Given that the coupled tensor $w^{abr}_{\rho}$ contains the important degrees of freedoms of the relevant electronic states, we express them in terms of the multipole basis.
The resulting form is
\begin{eqnarray}
X^{(b)}_{z\rho}(a-r)&=&i^{a+b-r}(-1)^{a}\sqrt{2[b,r]}n_{la}n_{1/2b}\underline{n}_{abr}
\notag\\&\quad&
\times \braket{l||X^{(\text{orb})}_{a}||l}w^{abr}_{\rho}(l),
\end{eqnarray}
where $\braket{l||X^{(\text{orb})}_{a}||l}$ is the reduced matrix element of the orbital multipole operator.
This relation reformulates the sum rule (2) in terms of the multipole basis $X_{lm}^{(s)}(k)$.
The explicit corresppondence between the representative multipoles and the coupled tensor is given in Table~\ref{tbl1}.

Equation (2) is then applied to the $L_3$ absorption edge of $3d$ T in TF$_2$.
In this case, the resonant emission process involves excitation $2p \rightarrow 3d$ followed by de-excitation $3s \rightarrow 2p$. The corresponding angular momenta are $c_1 = 1$, $j_1 = \frac{3}{2}$, $c_2 = 0$, and $j_2 = \frac{1}{2}$, yielding $L = 1$ and $L' = 1$.
Therefore, the summations run over $z = \{0, 1, 2\}$ and $z' = \{0, 1, 2\}$, respectively.
For this absorption edge, the sum of RXES integrated intensities for each polarization component, $I_{\mu}$, defines the total intensity $I_{\text{sum}} = I_{1} + I_{-1}$, while their difference defines as the RXES-MCD intensity $I_{\text{MCD}} = I_{1} - I_{-1}$. As shown in Fig.~1, there are two T sites, and the sum rule is applied to each site individually. The results are then obtained by evaluating the expectation value of both sites per unit cell.
The sum rule at each site ($i=1,2$) is expressed as
\begin{eqnarray}
I^{\text{RXES}}_{\text{sum},i}
&=&\frac{4}{\sqrt{3}}\mathbb{N}(211\omega)\tilde{\mathcal{B}}^{0}_{\frac{3}{2},\frac{1}{2}}(101)\Theta^{000}\left(\boldsymbol{\hat{k}},\boldsymbol{\hat{k'}}\right)
\notag\\
&\times&\left(2\braket{\mathbb{Q}_{0}}_i+\frac{\sqrt{3}}{2}\braket{\mathbb{Q}_{0}'}_i\right)
\notag\\
&-&\left[\frac{28}{15\sqrt{2}}\mathbb{N}(211\omega)\tilde{\mathcal{B}}^{2}_{\frac{3}{2},\frac{1}{2}}(101)\Theta^{022}\left(\boldsymbol{\hat{k}},\boldsymbol{\hat{k'}}\right)\right.
\notag\\
&+&\frac{28}{3\sqrt{2}}\mathbb{N}(211\omega)\tilde{\mathcal{B}}^{0}_{\frac{3}{2},\frac{1}{2}}(101)\Theta^{202}\left(\boldsymbol{\hat{k}},\boldsymbol{\hat{k'}}\right)
\notag\\
&+&\frac{14}{3}\mathbb{N}(211\omega)\tilde{\mathcal{B}}^{2}_{\frac{3}{2},\frac{1}{2}}(101)\Theta^{220}\left(\boldsymbol{\hat{k}},\boldsymbol{\hat{k'}}\right)
\notag\\
&+&\left.\frac{2\sqrt{5}}{3}\mathbb{N}(211\omega)\tilde{\mathcal{B}}^{2}_{\frac{3}{2},\frac{1}{2}}(101)\Theta^{222}\left(\boldsymbol{\hat{k}},\boldsymbol{\hat{k'}}\right)
\right]\braket{\mathbb{Q}_{u}}_i
\notag\\
&+&12\mathbb{N}(211\omega)\tilde{\mathcal{B}}^{2}_{\frac{3}{2},\frac{1}{2}}(101)\Theta^{224}\left(\boldsymbol{\hat{k}},\boldsymbol{\hat{k'}}\right)
\braket{\mathbb{Q}_{4a}}_i,
\end{eqnarray}
and
\begin{eqnarray}
&I^{\text{RXES}}_{\text{MCD},i}=\mathbb{N}(211\omega)\tilde{\mathcal{B}}^{0}_{\frac{3}{2},\frac{1}{2}}(101)\Theta^{101}\left(\boldsymbol{\hat{k}},\boldsymbol{\hat{k'}}\right)\notag\\
&\times\left(\frac{4}{3}\sqrt{\frac{1}{6}}\braket{\mathbb{M}_{z}'}_i-14\sqrt{\frac{5}{3}}\braket{\mathbb{M}_{z}''}_i\right)\notag\\&
-\mathbb{N}(211\omega)\tilde{\mathcal{B}}^{2}_{\frac{3}{2},\frac{1}{2}}(101)\Theta^{121}\left(\boldsymbol{\hat{k}},\boldsymbol{\hat{k'}}\right)\times\notag\\
&\left(\frac{2\sqrt{5}}{15}\braket{\mathbb{M}_{z}}_i+\frac{4\sqrt{5}}{9}\braket{\mathbb{M}_{z}'}_i-\frac{7\sqrt{2}}{9}\braket{\mathbb{M}_{z}''}_i\right)\notag\\&
+\frac{\sqrt{7}}{9}\mathbb{N}(211\omega)\tilde{\mathcal{B}}^{2}_{\frac{3}{2},\frac{1}{2}}(101)\Theta^{123}\left(\boldsymbol{\hat{k}},\boldsymbol{\hat{k'}}\right)\notag\\
&\times\left(\braket{\mathbb{M}_{xyz}'}_i+\braket{\mathbb{M}_{\alpha_z}'}_i\right).&
\end{eqnarray}
In (6) of the RXES sum rule, the terms include the number of holes $\braket{\mathbb{Q}_0}$, the spin-orbit interaction $\braket{\mathbb{Q}_0'}$, the electric quadrupole $\braket{\mathbb{Q}_u}$, and the electric hexadecapole $\braket{\mathbb{Q}_{4a}}$ as shown in Table~I.
On the other hand, in (7) of the RXES-MCD sum rule, in additionally involves the orbital magnetic moment $\braket{\mathbb{M}_z}$, the spin magnetic moment $\braket{\mathbb{M}_z'}$, and the anisotropic magnetic dipole $\braket{\mathbb{M}_z''}$ appear.
Importantly, magnetic octupoles $\braket{\mathbb{M}_{xyz}'}$ and $\braket{\mathbb{M}_{\alpha_z}'}$ also appear.

As shown in Fig.~1, when the $\ket{zx\uparrow}$ orbital is occupied at Site~1 and the $\ket{yz\downarrow}$ orbital is occupied at Site~2, its electronic configuration corresponds to the ferroic magnetic octupole order~\cite{SpaldinPRX2024}.
Expectation values of all the above operators are then taken with respect to these electronic configurations.

In the crystal coordinate system $x'y'z'$, the orbital state $\ket{zx \uparrow}$ at site~1 is re-expressed as $\tfrac{1}{\sqrt{2}}(\ket{z'x' \uparrow} - \ket{y'z' \uparrow})$, while the orbital state $\ket{yz \downarrow}$ at Site~2 becomes $\tfrac{1}{\sqrt{2}}(\ket{z'x' \downarrow} + \ket{y'z' \downarrow})$. Taking the expectation values with these states, the spin magnetic moments reflect their antiferromagnetic character, yielding $\braket{\mathbb{M}_z'}_1 = -\braket{\mathbb{M}_z'}_2$. The same sign reversal between Site~1 and Site~2 applies to the other two dipole components. By contrast, the magnetic octupole satisfies $\braket{\mathbb{M}_{xyz}'}_1 = \braket{\mathbb{M}_{xyz}'}_2$, reflecting the \textit{ferroic} nature of the order parameter.
Consequently, unit-cell summation cancels the first and second terms in (7), leaving only the third term. Furthermore, because $\braket{\mathbb{M}_{\alpha_z}'}_i$ and $\braket{\mathbb{M}_z'}_i$ belong to the same irreducible representation, it follows that $\braket{\mathbb{M}_{\alpha_z}'}_1=-\braket{\mathbb{M}_{\alpha_z}'}_2$. 
This indicates that, in rutile-type altermagnetic compounds, the expectation value of the magnetic octupole $\braket{\mathbb{M}_{xyz}'}$ $-$the essential order parameter$-$ can be extracted from the sum rule of RXES-MCD.

\renewcommand{\arraystretch}{1.0}
\begin{figure}[t]
  \begin{center}
   \includegraphics[width=70mm]{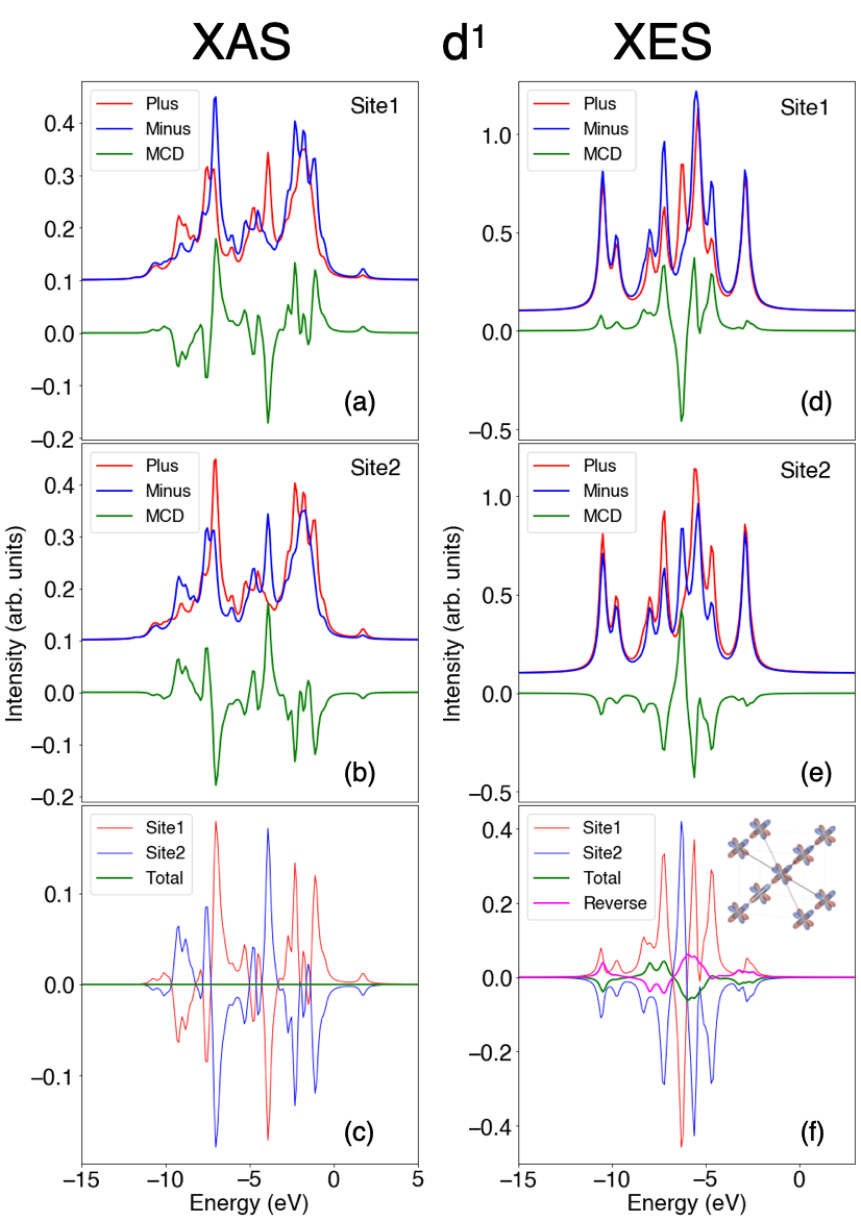}
  \caption{
  (Left) XAS-MCD spectra and (Right) RXES-MCD spectra of the $d^1$ configuration of TF$_2$. The RXES and MCD spectra are excited at $E=-6.8$ eV of XAS. (a, b) XAS and MCD spectra of T $L_{2,3}$-edges at Site~1 and at Site~2, respectively. (c) Sum of XAS-MCD spectra at both sites of T $L_{2,3}$-edges (green line). (d,e) RXES and MCD spectra at Site~1 and at Site~2, respectively. (f) Sum of RXES-MCD spectra at both sites (green line). The magenta line shows spectra for the flipped N\'{e}el vector.}
  \end{center}
\end{figure}
\begin{figure}[t]
  \begin{center}
   \includegraphics[width=70mm]{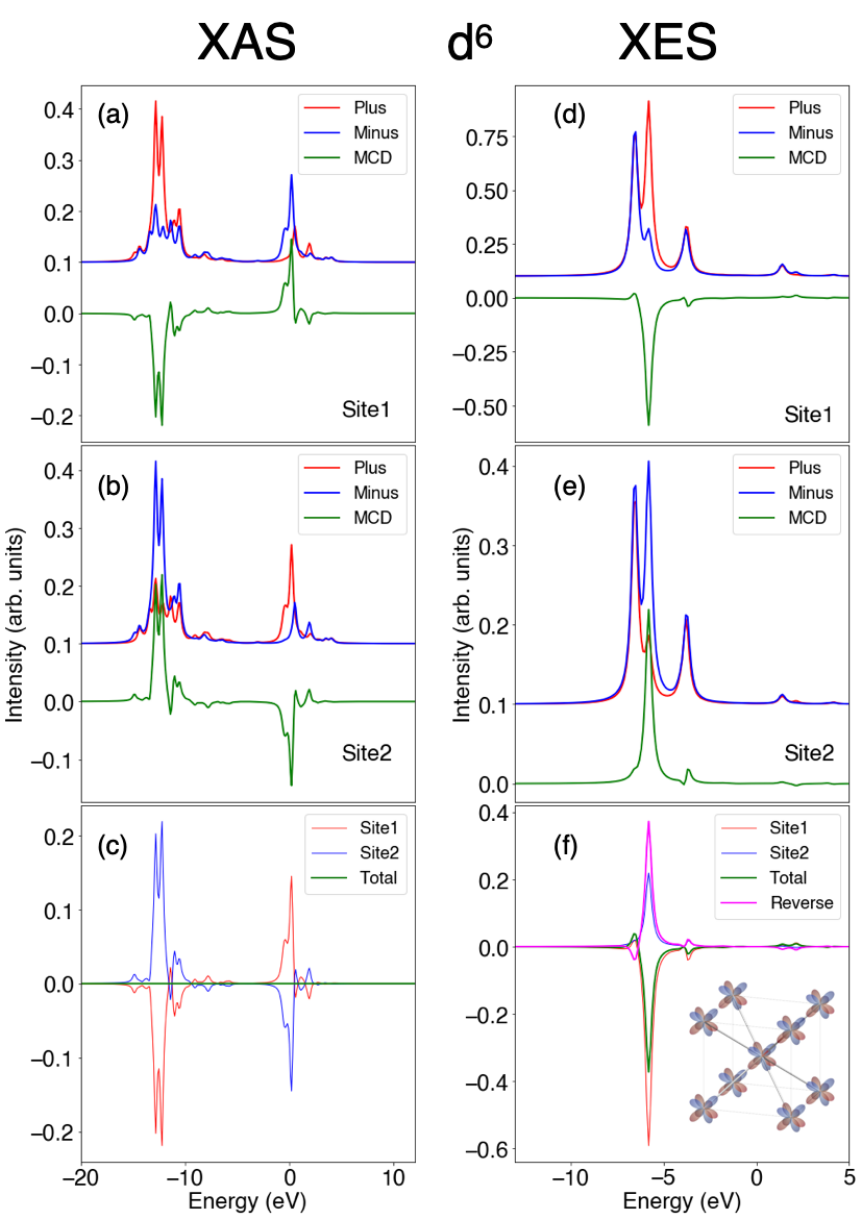}
  \caption{
    (Left) XAS-MCD spectra and (Right) RXES-MCD spectra of the $d^6$ configuration of TF$_2$. The RXES and MCD spectra are excited at $E=-12.8$ eV of XAS. (a, b) XAS and MCD spectra of T $L_{2,3}$-edges at Site~1 and at Site~2, respectively. (c) Sum of XAS-MCD spectra at both sites of T $L_{2,3}$-edges (green line). (d,e) RXES and MCD spectra at Site~1 and at Site~2, respectively. (f) Sum of RXES-MCD spectra at both sites (green line). The magenta line shows spectra for the flipped N\'{e}el vector.}
  \end{center}
\end{figure}

To verify this consideration, we calculate the $3s \rightarrow 2p$ emission spectra and their dichroism at the $L_3$ absorption edge of $3d$ T elements using an effective model that fully incorporates the multiplet structure~\cite{OkadaJPSJ,TanakaJPSJ,TaguchiJPSJ,MatsubaraJPSJ}. Calculations are performed for Site~1 and Site~2 in Fig.~1, and the spectra per unit cell are obtained by summing over their contributions. At each site, a crystal field and effective magnetic field of $D_{2h}$ symmetry are applied so that an electron occupies the $\ket{zx \uparrow}$ orbital at Site~1 and the $\ket{yz \downarrow}$ orbital at Site~2. The calculation parameters are given in the Supplemental Material~\cite{test}. 
Both single-electron ($d^1$) and multi-electrons ($d^6$) configurations are investigated. 

First, we show the spectra for the $3d^1$ configuration. Figures~2(a) and (b) display the XAS-MCD spectra for Site~1 and Site~2, respectively, while (c) shows the sum of the MCD spectra from both sites.
Meanwhile, Figures~2(d) and (e) show the RXES-MCD spectra for Site~1 and Site~2, respectively, and (f) shows their total contribution. As shown in Figs.~2(a) and (b), the XAS-MCD spectra have opposite signs at Site~1 and Site~2, reflecting their antiparallel magnetic moments. 
The sum of these XAS-MCD spectra (green line) vanishes over the full energy range in Fig.~2(c). Similarly, the RXES-MCD spectra exhibit opposite signs at the tow sites, as shown in Figs.~2(d) and (e). However, the sum of the RXES-MCD spectra remains finite as shown in Fig.~2(f) with an energy-integrated value of $-0.038$. According to the aforementioned sum rule, this integrated value originates solely from the magnetic octupole $\braket{\mathbb{M}_{xyz}'}$. Therefore, an experimental evaluation of the integrated intensity of the RXES-MCD spectrum would provide direct evidence of the ferroic ordering of magnetic octupole in rutile-type antiferromagnets. When N\'{e}el vector is flipped, the RXES-MCD spectrum (magenta line in Fig.~2(d)) reverses completely, reflecting the opposite sign of $\braket{\mathbb{M}_{xyz}'}$. 

Next, we examine the multi-electron $3d^6$ configuration ; the resulting XAS-MCD and RXES-MCD spectra are shown in Fig.~3, corresponding to Fig.~2. 
As in the $3d^1$ configuration, the XAS-MCD spectra for Site~1 and Site~2 exhibit opposite signs, reflecting the antiferromagnetic nature of their magnetic dipole moments as shown in Figs.~3(a) and (b). The sum of these XAS-MCD spectra (green line) shows in the vanishing intensity, as in Fig.~2(c). The energy-integrated value also vanishes within numerical accuracy, demonstrating that the MCD signal per unit cell cancels.
Similarly, the RXES-MCD spectra exhibit opposite signs between Site~1 and Site~2, as shown in Figs.~3(d) and (e), while their sum remains finite as shown in Fig.~3(f). The energy-integrated value is $-0.093$. As in the $3d^1$ configuration, this result indicates that ferroic ordering of magnetic octupole is detectable even in the many-electron configuration. As in $d^1$ configuration. When the N\'eel vector is flipped, the RXES-MCD spectrum is inverted.

In summary, we applied the sum rule for magnetic circular dichroism in the resonant X-ray emission process ($2p \rightarrow 3d : 3s \rightarrow 2p$) to rutile-type $3d$ compounds TF$_2$, which is a typical candidate of $d$-wave altermagnets, 
and demonstrated that the expectation value of the third-rank magnetic octupole is finite, reflecting the \textit{ferroic} nature of the essential order parameter.
Using the effective model with full multiplet structure, we further demonstrated that the total intensity remains finite in RXES-MCD, although the MCD intensity vanishes in XAS. The sign of the integrated intensity in RXES-MCD corresponds to that of the magnetic octupole.

This fact clearly indicates that RXES-MCD measurements provide a promising probe of the ferroic higher-order magnetic multipoles, and X-ray spectroscopy offers a powerful means to identify both the type and rank of active multipoles.
Thus, X-ray spectroscopy now offers rank and type selectivity of multipoles, in addition to its conventional element and orbital selectivity.
In the future, thourgh integrative analyses of X-ray absorption, emission, and photoemission spectroscopy data with the aid of information-science approaches such as Bayesian inference~\cite{YokoyamaJPSJ2021}, research on ferroic ordering will cover not only for magnetic multipoles but also for electric, electric-toroidal, and magnetic-toroidal multipoles.

\begin{acknowledgments}
We thank Drs. Y. Yamasaki and A. Hariki for fruitful discussion.
This work was supported by JSPS KAKENHI Grants Numbers JP23K03288 and JP23H00486.
\end{acknowledgments}

\bibliographystyle{apsrev4-2}
\bibliography{20250918_Detection_of_ferroic_octupole_ordering_in_d_wave_Altermagnetic_rutile_type_compounds_HK.bbl}

\end{document}